\documentclass[referee]{aa} 
\usepackage{graphics}
\begin{document}

   \thesaurus{03     
              (03.09.2;  
               03.19.2  )} 
   \title{Present and future gamma-ray burst experiments}

   \author{K. Hurley}

   \offprints{K. Hurley}

   \institute{University of California at Berkeley, Space Sciences Laboratory,
              Berkeley, CA, USA 94720-7450\\
              email: khurley@sunspot.ssl.berkeley.edu}

   \date{Received ; accepted }

   \maketitle

   \begin{abstract}

Gamma-ray burst counterpart studies require small, prompt error boxes.  Today,
there are several missions which can provide them: BeppoSAX, the Rossi X-Ray
Timing Explorer, and the 3rd Interplanetary Network.  In the near future, HETE-II,
a possible extended Interplanetary Network, and INTEGRAL will operate in this
capacity.  In the longer term future, a dedicated gamma-ray burst MIDEX mission
may fly.  The capabilities of these missions are reviewed, comparing the number
of bursts, the rapidity of the localizations, and the error box sizes.

   \end{abstract}

%

\section{Introduction}

   The long-awaited breaktkhrough in our understanding of cosmic gamma-ray
bursts (GRBs) has come about because accurate ($<$10 \arcmin) burst positions have become
available quickly ($<$1 day).  Prior to the launch of BeppoSAX, accurate
positions were available from the interplanetary networks, but unavoidable
delays in the retrieval and processing of data delayed their availability.
Similarly, rapidly determined positions were, and still are available from BATSE,
but their utility is limited by the fact that their accuracy is in the
several-degree range.  

The list of things we need to know about bursts is still long.  Among the items
on it are:

\begin{itemize}
\item{are burst sources in their host galaxies, or outside them?}
\item{what is the distribution of GRB distances?}
\item{are bursts beamed?}
\item{what is the intrinsic luminosity function for bursts?}
\item{are there different classes of bursts, e.g. long and short,
	 soft-spectrum and hard-spectrum?}
\item{what is the multiwavelength behavior of GRB light curves immediately
	 after the burst?}
\end{itemize}

Given that only $\sim 50\%$ of the GRBs studied to date have optical counterparts,
and that only $\sim 50\%$ of the counterparts have measured redshifts, it is clear that
answering these questions will require hundreds of GRB detections in the
gamma-ray range.  But the rate at which rapid, accurate positions become available is still quite
small: $<$ 1 burst/month.  Thus even minor improvements in the rate can
have a major impact on progress in the near-term future.  However, major
improvements will be needed in the long-term future to make the next big step.


\section{Current and future missions}

Figure 1 shows the approximate operating dates of the missions which are
capable of providing GRB data to answer some of the questions listed above.
Each of these missions will be reviewed briefly.

\begin{figure*}
   \resizebox{\hsize}{!}{\includegraphics{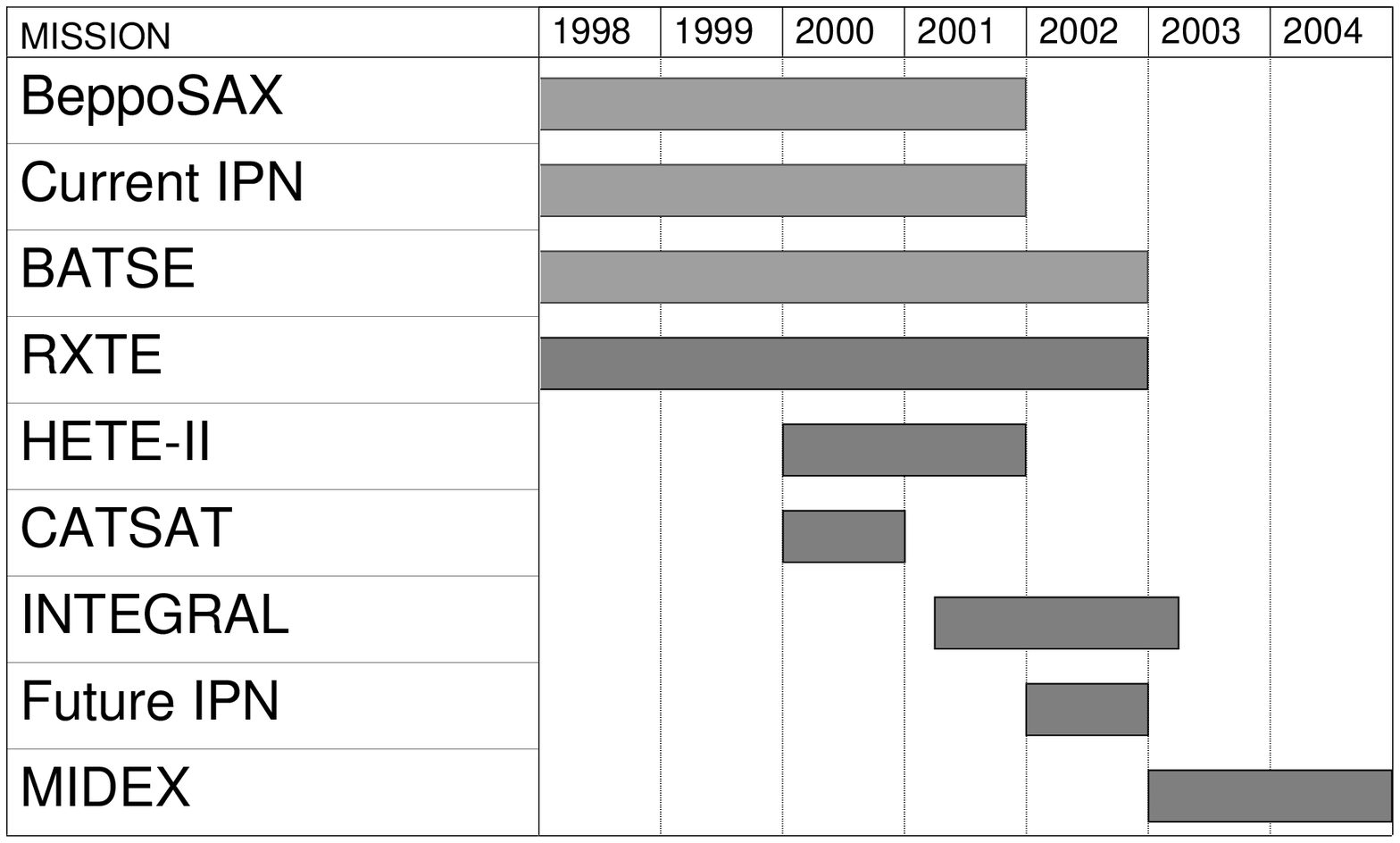}}
   \caption{Approximate dates of current and future GRB missions}
   \label{Figure 1}
\end{figure*}

\subsection{BeppoSAX}

BeppoSAX has now observed 14 GRBs in the Wide Field Camera, with
location accuracies in the $<10 \arcmin$ range.  The resulting detection
rate is $\sim$ 8/year.  Eleven of these have been
followed up with Narrow Field Instrument observations, resulting in
many cases in a reduction of the error
circle radii to $\sim 1 \arcmin$ (Costa \cite{costa}).  There are
delays of $\sim$ hours to obtain, analyze, and distribute the data.  The
approved lifetime of the mission is through 2001.

\subsection{BATSE: GCN and Locburst}

The Global Coordinates Network (GCN: Barthelmy et al. \cite{barthelmy})
distributes $\sim$ 300 GRB positions/year with delays of the order of seconds,
determined directly onboard the CGRO spacecraft.  The error circle radii
are $> 4 \degr$.    The Locburst procedure (Kippen et al. 
(\cite{kippen}) distributes $\sim$ 100 of the stronger bursts/year. 
As Locburst relies on ground-based processing, the delays are longer, $\sim$ 15 min,
but the accuracy is improved: the error circle radii are $>1.6 \degr$.  
These data are useful for follow-up searches with rapidly moving
telescopes like LOTIS (Park \cite{park}), and with the RXTE
Proportional Counter Array, as well as for triangulation
with the 3rd Interplanetary Network.  BATSE will remain operational at least through
2002; its lifetime is limited by the available funding, and is reviewed every
two years in NASA's ``Senior Review'' process.

\subsection{3rd Interplanetary Network}

The 3rd IPN consists of the Ulysses and Near Earth Asteroid Rendezvous (NEAR) missions
in interplanetary space, as well as numerous near-earth missions such as CGRO, RXTE, 
Wind, and BeppoSAX.  The IPN observes and localizes $\sim$ 70 GRBs/year (Hurley 
\cite{hurley1} a,b).  When a burst is observed by just two spacecraft,
such as Ulysses and CGRO, the resulting error box is the intersection of the triangulation
annulus with the BATSE error circle, with dimensions typically 5 $\arcmin$ by 5 $\degr$.
When Ulysses, NEAR, and say, BATSE detect the burst, the resulting error box may be as
small as 1 $\arcmin$ by 5 $\arcmin$ (Cline \cite{cline}).  The delays involved
are $\sim$ 1 day, imposed by the receipt of data from interplanetary spacecraft through
NASA's Deep Space Network.  The lifetime of the 3rd IPN will be through 2001 at least.
This is determined by the nominal end of the Ulysses mission, which will be reconsidered in
NASA's Senior Review of space physics missions in 1999.

\subsection{The Rossi X-Ray Timing Explorer}

The All-Sky Monitor aboard RXTE detects $\sim$ 5 GRBs/year; $\sim$2-3 of them
can be localized to $\sim$ arcminute accuracy with delays of only minutes (Bradt
\cite{bradt}).  In addition, the PCA performs about one target-of-opportunity
observation per month of BATSE Locburst positions to search for fading X-ray counterparts
(Takeshima et al. \cite{takeshima}).  When successful, the counterpart position
can be determined to $\sim 10 \arcmin$ with a delay of hours.  Like BATSE, RXTE's
lifetime, determined by the Senior Review, will extend through 2002 at least.

\section{The next big step}

At present we rely on spacecraft instrumentation to provide X-ray positions
which are accurate to arcminutes, and on rapid ground-based photometry from
small to moderate-sized telescopes to identify optical counterparts to arcsecond
accuracy.  Only at that point can a large telescope be used to determine the
redshift (for example, the spectrometer slits on the Keck Low Resolution Imaging
Spectrometer are only 1 - 8 $\arcsec$).  The next big step will be to determine
GRB positions directly on the spacecraft to arcsecond accuracy, eliminating the delays
involved in refining the positions on the ground.  Some of the future missions 
discussed below will be capable of accomplishing this.

\subsection{HETE-II}

The High Energy Transient Explorer-II combines a Wide Field X-ray Monitor
and a Soft X-ray Camera to localize $\sim$50 GRBs/y to accuracies of 10 $\arcsec$
to 5 $\arcmin$ (Ricker \cite{ricker}).  Locations will be transmitted to
the ground in near real-time.  The HETE-II mission is planned for a two year lifetime
starting in late 1999.

\subsection{CATSAT}

The Cooperative Astrophysics and Technology Satellite (Forrest et al. \cite{forrest})
will contain a soft X-ray spectrometer consisting of 190 cm$^2$ of Si avalanche
photodiodes to measure the 0.5 - 20 keV spectra of GRBs and their afterglows.
From these spectral
measurements, the hydrogen column along the line of sight may be determined.
CATSAT has only coarse localization capability, but measurements of N$_H$ will
help to answer the question of the locations of GRBs with respect to their
host galaxies.  $\sim 12$ GRBs/year should be detected, with data available
$\sim$ 5 hours after the bursts.  A nominal one year mission in 2000 is
planned.

\subsection{INTEGRAL}

The International Gamma-Ray Laboratory can detect bursts with its Ge spectrometer
array (the SPI), as well as with IBIS (the Imager on-Board the INTEGRAL Satellite), and
with the BGO anticoincidence shield around the spectrometer.  IBIS, a CdTe
array with a coded mask, provides the most accurate, rapid locations.
It can detect $\sim$ 20 GRBs/year and localize them to
arcminute accuracy (Kretschmar et al. \cite{kretschmar}).  These positions
can be distributed to observers within 5 - 100 s.  The nominal INTEGRAL mission
is two years long, starting in April 2001.

\subsection{Future IPN}

A future Interplanetary Network, consisting of Mars Surveyor Orbiter 2001, the
Near Earth Asteroid Prospector, INTEGRAL, and possibly BATSE and Ulysses, may
exist around the year 2002.  MSO has two GRB two instruments which will detect
GRBs with good sensitivity and time resolution, a Ge spectrometer
and a neutron detector.  The BGO anticoincidence shield of the INTEGRAL SPI is
similarly equipped to detect bursts (Hurley \cite{hurley3}).  A small
GRB detector has been proposed as a MIDEX mission of opportunity for NEAP, a
mission to Nereus.  With such a network, $\sim$ 70 GRBs/year could be localized
to arcminute accuracies, with delays of the order of a day.  This IPN might
remain in place for one or two years, bridging the gap to a possible dedicated
GRB MIDEX.

\subsection{A dedicated MIDEX}

Approximately 6 proposals for dedicated GRB missions were submitted to NASA
in response to the recent MIDEX announcement.  A dedicated MIDEX could localize
perhaps 100 GRBs/year to arcsecond accuracy onboard the spacecraft, and transmit
the locations to the ground in near real-time.  Such a mission may fly in the
years 2003-2005.  The announcement of the MIDEX selection is expected in
January 1999.

\section{Conclusions}

   \begin{table*}[t]
      \caption[]{Capabilities of current and future GRB missions}
      \label{tabone}
      \begin{tabular}{lllll} \hline
Mission           & Bursts/yr & Accuracy            & Delay        & Dates        \\ \hline
BeppoSAX          & 8         & 1 - 10 $\arcmin$    & hours        & through 2001 \\
BATSE (GCN)       & 300       & 8 - 20 $\degr$ dia. & seconds      & through 2002 \\
BATSE (Locburst)  & 100       & 3 - 8 $\degr$ dia.  & 15 - 30 min. & through 2002 \\
Current IPN       & 70        & $\rm5 \arcmin x 5 \degr - 1 \arcmin x 5 \arcmin $ & $\sim$ day &
through 2001 \\
RXTE ASM AND PCA  & 5         & 3 - 10 $\arcmin$    & min. - hours & through 2002 \\
HETE-II           & 50        & 10 $\arcsec$ - 5 $\arcmin$ & seconds & 2000 - 2002 \\  
CATSAT            & 12        & degrees             & 5 hours      & 2000 \\
INTEGRAL          & 20        & arcminutes          & 5 - 100 s    & 2001 - 2003 \\
Future IPN        & 70        & arcminutes          & $\sim$ day   & 2002 - 2003 \\
MIDEX             & 100       & arcseconds          & seconds      & 2003 - 2005 \\ \hline
     \end{tabular}
      \vspace{4in}
   \end{table*}

Table 1 summarizes the essential characteristics of the current and future
GRB missions.  Although we have much to learn about gamma-ray bursts, these
missions promise to return the data we need to move forward in this exciting
field.

\end{document}